\documentclass[aps,prb,twocolumn]{revtex4}

\usepackage[utf8]{inputenc} 

\usepackage[english,french]{babel} 
\usepackage[T1]{fontenc} 
\usepackage{lmodern} 

\usepackage{graphicx}
\usepackage{dcolumn}
\usepackage{amsmath}
\usepackage{color}
\usepackage{soul}
\usepackage{soulutf8}
\usepackage{ulem}
\usepackage{gensymb}
\usepackage{xcolor} 

\newcommand{\NCS}{$\kappa$-(ET)$_2$Cu(NCS)$_2$}

\newcommand\Tstrut{\rule{0pt}{2.6ex}}

\begin{document}

\title{Possible observation of the signature of the bad metal phase and its crossover to a Fermi liquid in $\kappa$-(BEDT-TTF)$_2$Cu(NCS)$_2$ bulk and nanoparticles by Raman scattering}

\author{M. Revelli Beaumont$^{1,2}$}
\author{P. Hemme$^{1}$}
\author{Y. Gallais$^1$}
\author{A. Sacuto$^1$}
\author{K. Jacob$^{2}$}
\author{L. Valade$^{2}$}
\author{D. de Caro$^{2}$}
\author{C. Faulmann$^2$} 
\author{M. Cazayous$^{1}$}
\affiliation{$^1$Laboratoire Mat\'eriaux et Ph\'enom\`enes Quantiques (UMR 7162 CNRS), 
Universit\'e de Paris, 75205 Paris Cedex 13, France\linebreak
$^2$Laboratoire de Chimie de Coordination (UPR 8241), Universit\'e Paul Sabatier Toulouse, France}

\begin{abstract}
$\kappa$-(BEDT-TTF)$_2$Cu(NCS)$_2$ has been investigated by Raman scattering in both bulk and nanoparticle compounds. Phonon modes from 20 to 1600 cm$^{-1}$ have been assigned. Focusing on the unexplored low frequency phonons, a plateau in frequencies is observed in the bulk phonons between 50 and 100 K and assigned to the signature of the bad metal phase. Nanoparticles of $\kappa$-(BEDT-TTF)$_2$Cu(NCS)$_2$ exhibit anomalies at 50 K associated to the crossover from a bad metal to a Fermi liquid whose origins are discussed.
\end{abstract}

\maketitle

\section{\label{sec:intro}Introduction}

Organic compounds are usually electrical insulators. However, charge-transfer complexes and salts, in which the large intermolecular $\pi$-orbital overlaps create a conduction pathway. Organic conductors have been thus a focus in the development of innovative architectures based on emerging molecular devices.\cite{Jalabert} From a fundamental point of view, organic compounds also present phases often in competition that have attracted attention.\cite{Dressel2018, Ishuguro1998, Lang2008, Powell2006, lebed2006, Drichko2015} Organic superconductors share various similarities with the phase diagram of high temperature copper oxide superconductors.\cite{Keimer1992}
Among the charge transfer salts, $\kappa$-(BEDT-TTF)$_2$Cu(NCS)$_2$, with BEDT-TTF [bis(ethylene,dithio)tetrathiafulvalene] molecules commonly abbreviated as ET, is a well known organic compound due to the occurence in its phase diagram of unconventional superconductivity close to an antiferromagnetic phase.\cite{Ishuguro1998, Lang2008, Powell2006, lebed2006} 

\begin{figure}[h!]
	\begin{center}
		\includegraphics[width=8.7cm]{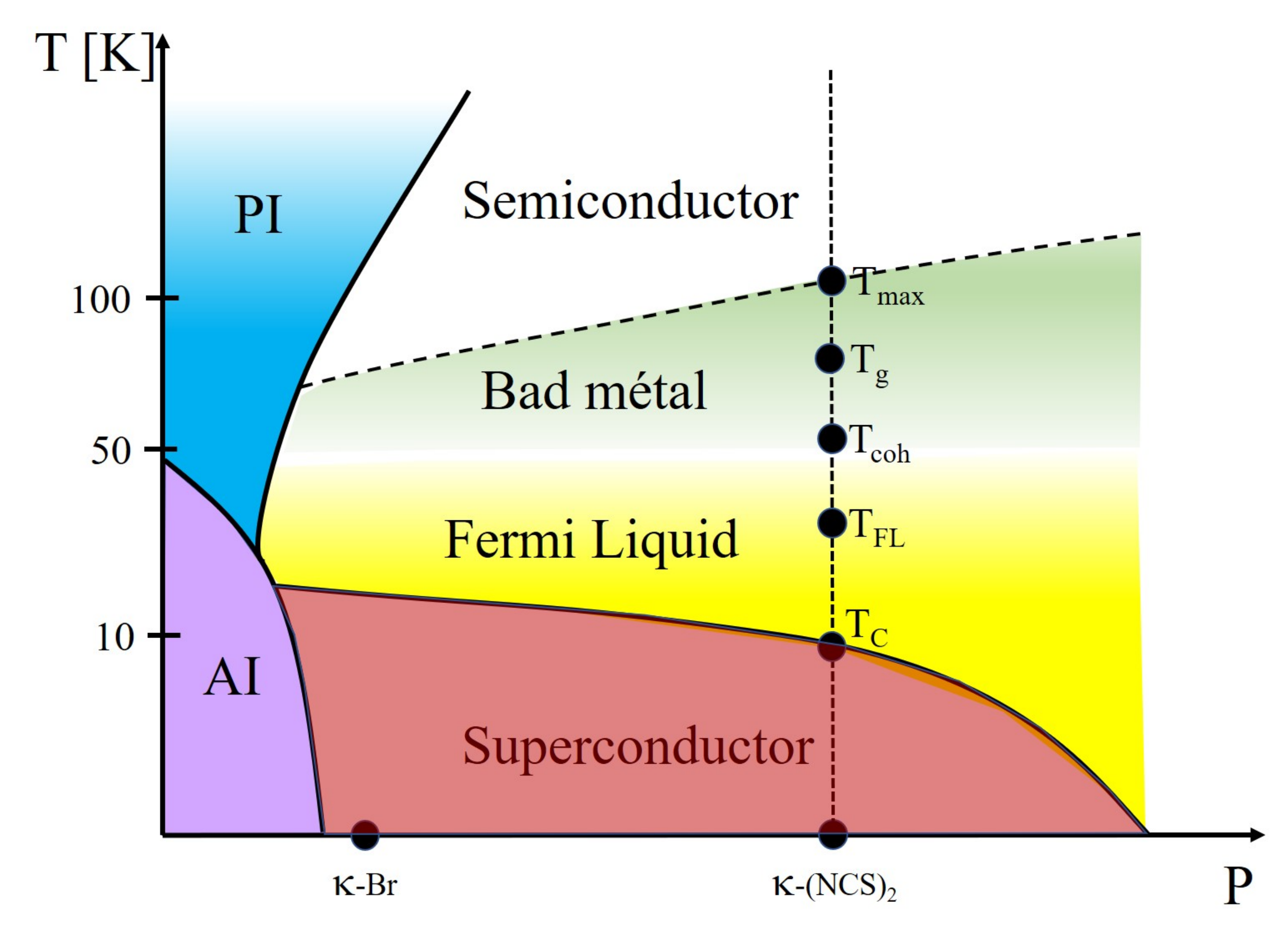}
		\caption{Schematic phase diagram of organic charge transfer salts as a function of chemical pressure (or external) for $\kappa$-(BEDT-TTF)$_2$Br and \NCS{}. PI and AI correspond to paramagnetic insulator and antiferromagnetic insulator phases, respectively.}
		\label{Fig0}
	\end{center}
\end{figure}

Indeed, \NCS{} exhibits a complex phase diagram at ambient pressure (see Fig.~\ref{Fig0}). 
\NCS{} is superconducting up to the critical temperature $T_C$ = 10.4 K.\cite{urayamaNewAmbientPressure1988, Sugano} 
Between $T_C$ and $T_{FL}$ = 20 - 25 K, the compound 
behaves like a Fermi liquid\cite{Dressel2007,Merino2008} characterized by a quadratic temperature dependence in the resistivity.\cite{Limelette2003,Strack2005, Milbrandt2013} 
Between $T_{FL}$  and $T_{coh}\simeq50$ K, the resistivity has no more a quadratic dependence, indicating the breakdown of true Fermi liquid behaviour.\cite{Limelette2003,Milbrandt2013,Georges2013} 
However, the charge transport is still dominated by coherent quasiparticles.\cite{Merino2008,Dressel2004,Frikach}
As the temperature is further raised above $T_{coh}$ the resistivity continues to increase monotonically until a broad maximum is reached at $T_{max}\simeq100$\,K.
The spectral weight at the Fermi energy is negligible \cite{Analytis2006,Su1998,Strack2005,Merino2000,Limelette2003} and the resistivity starts to decrease monotonically. The electron-electron interactions control this phase as shown by dynamical mean-field theory (DMFT).\cite{Limelette2003,Merino2000,Georges2013,Kotliar2004} This phase is described by a bad metal as an intermediate temperature state of a metal. 
It is characterized for example by ill-defined quasi-particles.
Ultrasonic attenuation experiments show that the electron-phonon coupling is strong at the crossover between the two phases.\cite{Frikach} The vibrational degrees of freedom might thus be used as a signature of this transition.
The electrical resistance of \NCS  shows semiconductor behavior above $T_{max}$.\cite{Kuwata} 
The Fermi-surface topology has also been investigated and determined by magnetotransport experiments.\cite{Singleton, Caulfield}

The spin degrees of freedom are strongly impacted with temperatures.
Above $T_{coh} \simeq$ 50 K, the spin correlations are captured by a phenomenological spin-fluctuation model from Millis, Monien, and Pines \cite{Moriya,MMP,Yusuf1,Kawamoto95,Kawamoto95B} which quantitatively reproduces the monotonic increase in spin correlations observed in NMR experiments.  
DMFT calculations fail to depict the NMR experiment results below this temperature (strong decrease of the Knight shift, for example).\cite{Yusuf1} Three hypotheses have been proposed to explain the NMR measurements below $T_{coh}$ : the loss of the antiferromagnetic spin correlations below 50 K, the opening of a pseudogap or the opening of a real gap associated with $T_{coh}$ on the small 1D-parts of the FS as opposed to a pseudogap on the major quasi-2D fractions.\cite{Powell2011} However, the pseudogap opening has only been observed up to now in NMR experiments.

Note that the ET molecules are also involved in the physics of the metallic, insulating and superconducting phases.\cite{Muller2002,Scriven2009b,Powell2004,Taniguchi2003} The conformation of the ET molecules that can be staggered or eclipsed induces a glassy transition around $T_g$ = 80 K.  Above $T_g$ there is a thermal distribution of staggered and eclipsed states whereas below $T_g$ the dynamics of the terminal ethylene groups are strongly suppressed and a order-disorder transition is observed.\cite{Muller2002} 

Most of the previous Raman scattering measurements on \NCS{} have been devoted to the study of the superconducting phase.\cite{Dressel1992, Pedron1997,Lin2001,Truong2007} In particular, anomalies in the Raman shift and in the broadening of the phonon modes have been interpreted as the signature of the pair breaking energy.\cite{Zeyher1990} In this work, we bring insight into the rich physics of \NCS{} through the study of the intermolecular phonon modes below 200 cm$^{-1}$ which give access to the electron correlations and intermolecular magnetic coupling. We have implemented this approach in \NCS{} bulk and nanoparticles (NPs) in order to study the effects of the size reduction on the \NCS{} properties. 
In the low frequency phonons of the bulk, we point out the signature of the transition from a semiconductor behavior to the bad metal. For NPs, additional anomalies have been observed at as a function of the temperature around $T_\textrm{coh}$ $\simeq$ 50 K and associated to the crossover from the bad metal to the Fermi liquid. 

\section{\label{sec:expDetails}Experimental Details}


High quality \NCS{} single crystals have been grown by electrocrystallization following the synthesis described in Ref.~{\onlinecite{urayamaNewAmbientPressure1988} and characterized by X-ray crystallography and resistivity measurements. Single-crystalline \NCS{} NPs with a mean diameter of 28 nm and a standard deviation of 4 nm, have been synthesized by chemical route using a biobased amphiphilic molecule, the dodecanoic acid C$_{11}$H$_{23}$COOH as the growth controlling agent. Their quality has been characterized by X-ray crystallography and IR measurements. TEM measurements evidence roughly spherical NPs.\cite{revellibeaumontReproducibleNanostructurationSuperconducting2020} Note that NPs differ from powder by  well defined and regular shapes and a controlled mean diameter with a small standard deviation.

The structure of mixed-valence salt  $\kappa-\text{(ET)}_2X$ consists of alternating layers of acceptors X and electron donor ET molecules. ET layers consist of pairs of ET molecules, which stacks almost perpendicular to each other, the long axis of the molecule always pointing in the same direction. This forms then a two dimensional conducting layer in the bc-plan,\cite{Lang2008, Dressel2018, Powell2006} which alternates with the insulating layer of the polymeric anion along the a-direction.
The ET molecule pairs donate an electron to the anion layer, leaving a hole in the highest occupied molecular orbital (HOMO) of ET. The $\kappa$-phases belong to the family of organic quasi-two-dimensional conductors with an in-plane conductivity much larger than the out-of-plane conductivity.

The Raman spectroscopy measurements were recorded using a triple substractive T-64000 Jobin-Yvon spectrometer with a cooled CCD detector. The spectra were acquired with a 532 nm laser line (green light) from an Oxxius-Slim solid state laser, filtered both spatially and in frequency and obtained in quasi-backscattering geometry. Measurements between 15 and 300 K have been performed using an ARS closed-cycle He cryostat. In order to avoid heating and damaging of the sample, the incident laser beam power was kept at 2 mW with a spot size of 100 $\mu$m diameter. Phonon modes were analyzed using a Lorentzian lineshape. The resolution of the phonon frequencies was around 0.2 cm$^{-1}$.

\section{\label{sec:Results}Results and discussion}

\subsection{\label{sec:IntramolVib}Phonon modes in the high frequency region}

Phonons in organic conductor salts are generally analyzed on the basis of the molecular vibrations in the neutral molecule. For \NCS{}, observed vibrational peaks are assigned to the ET molecule.

Neglecting the staggered/eclipsed distortion, the ET molecules can be supposed to be flat except for the hydrogen atoms and belong to the D$_{2h}$ point group. It has 72 possible phonon modes, of which 36 are Raman active\cite{kozlovAssignmentFundamentalVibrations1987}
\begin{align}
    \Gamma(D_2h) = 12 a_g + 6 b_{1g} + 7 b_{2g} + 11 b_{3g} + \\
    \nonumber
    7 a_u + 11 b_{1u} + 11 b_{2u} + 7 b_{3u}
\end{align}

However, ab initio quantum chemical calculations show that ET is non planar\cite{demiralpPredictionNewDonors1995, demiralpElectrontransferBoatvibrationMechanism1995}. The stable boat structure of ET has C$_2$ symmetry, leading to the mode distribution 
\begin{align}
    \Gamma(C_2) = 37 a + 35 b
\end{align}

The reduction of symmetry is as follows\cite{demiralpVibrationalAnalysisIsotope1998} : 
\begin{subequations}
    \begin{align}
        12 a_g + 7 a_u + 11 b_{3g} + 7 b_{3u} \rightarrow 37a \\
        11 b_{1u} + 6 b_{1g} + 11 b_{2u} + 7 b_{2g} \rightarrow 35b
    \end{align}
\end{subequations}

As a consequence, all the vibrational modes are Raman as well as infrared active due to this low-symmetry structure (the unit cell does not possess inversion symmetry).

There have been several works devoted to assign the modes associated to the neutral ET molecule.\cite{Eldridge1995, Lin1999}
Recently, ab-initio calculations of the vibrational properties of some compounds from this series reveal that a simple distinction between molecular vibrations coupling the dimers and lattice vibrations cannot be sustained.\cite{Dressel2016a, Dressel2016b} This implies that the simple assignment of pure phonon modes is in principle incorrect. However, nowadays, there is no calculation of phonon modes frequencies taking into account all these interactions. 

An assignment of high energy modes can be found in Refs \onlinecite{Eldridge1997} and \onlinecite{Eldridge2002}. 
In addition, Sugai {\it et al.} \cite{sugaiRamanactiveMolecularVibrations1993} assigned \NCS{} experimental vibrational modes using calculations made by Kozlov {\it et al.} \cite{kozlovAssignmentFundamentalVibrations1987} on neutral molecule ET$^0$ vibrations. 
On the other hand for IR measurements, Kornelsen {\it et al.} \cite{kornelsenInfraredOpticalProperties1991} assigned \NCS{} experimental vibrational modes using calculations made by Kozlov {\it et al.} \cite{kozlovElectronMolecularVibration1989} on cation ET$^+$ vibrations.

Nevertheless, in \NCS{} the charge redistribution on the ET atoms causes a shift of vibrational energy.\cite{sugaiRamanactiveMolecularVibrations1993} The total charge per dimer is thus neither ET$^0$ nor ET$^+$ but ET$^{0.5+}$. Hence, we propose to assign ET vibration modes with respect to an averaged calculated value of ET$^0$ and ET$^+$.


\begin{figure}[h]
	\begin{center}
		\includegraphics[width=8.7cm]{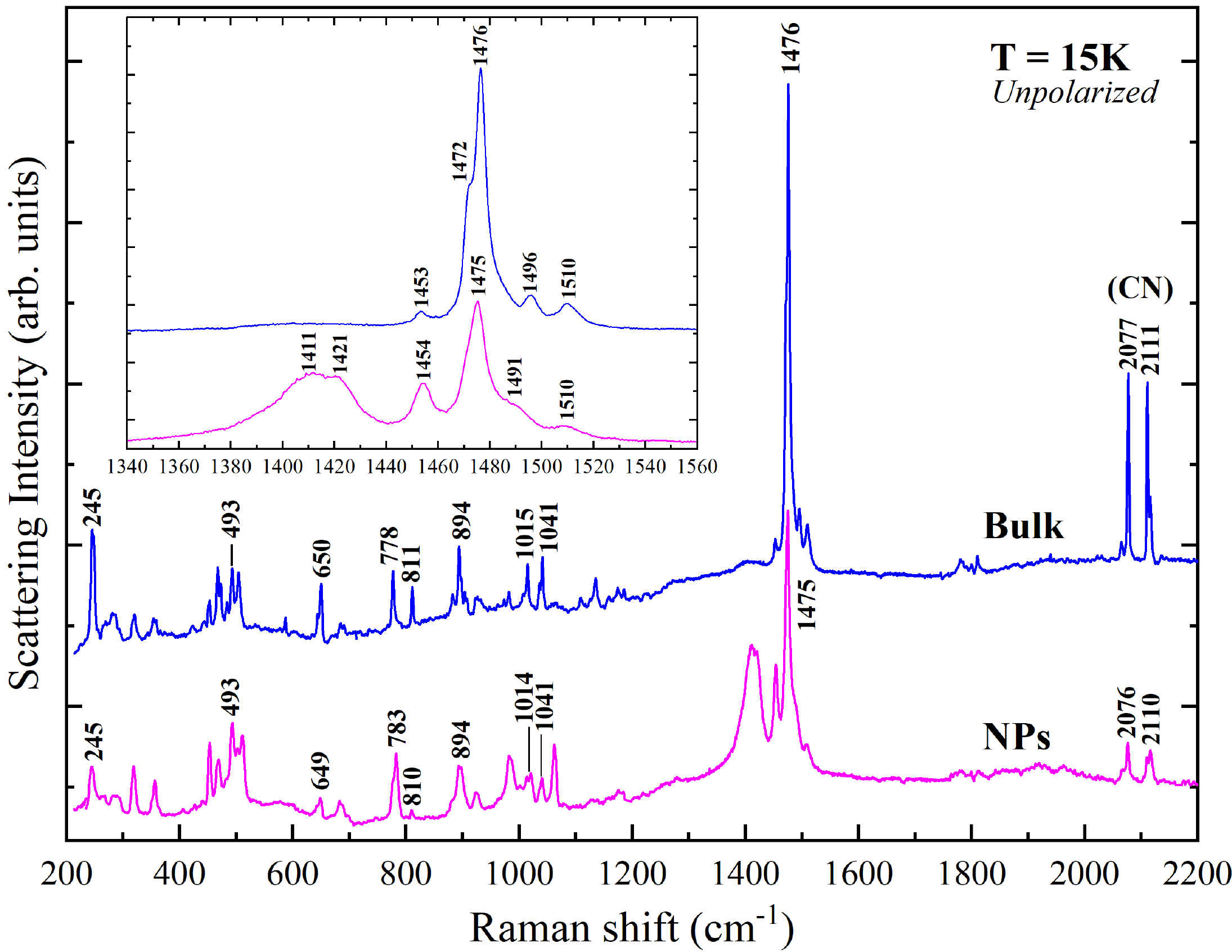}
		\caption{Unpolarized Raman spectra of NPs and bulk of \NCS{} at 15 K in the range of 200 – 2200 cm$^{-1}$. Inset : Zoom in the range of 1340 – 1560 cm$^{-1}$.}
		\label{Fig1}
	\end{center}
\end{figure}

Figure~\ref{Fig1} presents unpolarized Raman spectra at 15 K from 200 cm$^{-1}$ to 2200 cm$^{-1}$ of \NCS{} for bulk and NPs. Despite intensity variations and
broadening of phonon peaks, there are no main differences in peaks positions between the NPs and the bulk. Since NPs are randomly oriented, there are no Raman selection rules. Phonons in NPs are an averaged response of all possible bulk orientations.

\begin{table}    
    \begin{ruledtabular}
       \begin{tabular}{ccc|cccc}
            \multicolumn{3}{c|}{Assignment} & \multicolumn{4}{c}{Experimental}\\
            \hline
            $\nu_i$ & Symm & Calc. & \multicolumn{2}{c}{295K} & \multicolumn{2}{c}{15K} \Tstrut \\
            \cline{4-5} \cline{6-7}
            & & ET$^{0.5+}$ & Bulk & NPs & Bulk & NPs \\
            \hline
            2  & a (a$_g$)    & 1508 & 1501 & 1501 & 1510 & 1510 \Tstrut \\ 
            27 & b (b$_{1u}$) & 1478 & 1484 & 1484 & 1496 & 1491         \\
            3  & a (a$_g$)    & 1460 & 1467 & 1466 & 1476 & 1475         \\
               &              &      &      &      & 1472 &              \\
               &              &      &      & 1446 & 1453 & 1454         \\
            4  & a (a$_g$)    & 1422 &      & 1413 &      & 1421         \\
               &              & 1409 &      & 1402 &      & 1411         \\ 
            ?  &              &      &      & 1053 &      & 1063         \\
            58 & a (b$_{3g}$) & 1035 & 1031 & 1030 & 1041 & 1041         \\
            47 & b (b$_{2u}$) & 1011 & 1007 & 1010 & 1015 & 1014         \\
            6  & a (a$_g$)    & 985  & 972  & 979  & 982  & 982          \\
            7  & a (a$_g$)    & 899  & 893 & 893 & 894 & 894        \\
               & anion        &      & 806  &      & 811 & 810    \\
            61 & a (b$_{3g}$) & 782  & 772  & 773  & 778 & 783           \\
            8  & a (a$_g$)    & 649  & 642  &      & 650 & 649           \\
            9  & a (a$_g$)    & 498  & 500  & 502  & 504 & 502           \\
            34 & b (b$_{1u}$) & 501  & 483  & 483  & 493 & 493           \\
            10 & a (a$_g$)    & 462  & 466  & 462  & 467 & 468           \\
               & anion        &      & 445  & 446  & 453 & 453    \\
            63 & a (b$_{3g}$) & 351  & 354  & 352  & 354 & 356           \\
            11 & a (a$_g$)    & 314  & 311  & 310  & 317 & 318           \\
            53 & b (b$_{2u}$) & 263  & 262  & 262  & 267 & 267           \\
               & anion        &      & 244  & 244  & 245 & 245           \\
        \end{tabular}
    \end{ruledtabular}
    \caption{Experimental frequencies (in cm$^{-1}$) of the intense phonons from Fig.~\ref{Fig1}. Calculated frequencies for ET$^{0.5+}$, corresponding symmetry and mode labels are also reported.}
    \label{Table1}
\end{table}

In Tab.~\ref{Table1} are reported several phonon frequencies extracted from Fig.~\ref{Fig1} measured in bulk and NPs ambient temperature and 15 K as well as their corresponding symmetry and calculated frequencies for ET$^{0.5+}$. The latter averaged was derived using calculations on ET$^{0}$ and ET$^{+}$. \cite{demiralpVibrationalAnalysisIsotope1998, kozlovElectronMolecularVibration1989}
Experimental frequencies at room temperature agree correctly with modes of ET$^{0.5+}$. Note that the 806 and 445 cm$^{-1}$ modes are assigned to the anion modes.\cite{sugaiRamanactiveMolecularVibrations1993} 
At 15 K, additional modes are measured as the vibrations at 1472 and 1453 cm$^{-1}$ in the bulk and the vibration at 1411 cm$^{-1}$ in the NPs. Additional modes could have two origins, the splitting or degeneracy of predicted modes. Here, these modes might come from the $\nu_3$ and $\nu_4$ modes at 1460 cm$^{-1}$ and 1422 cm$^{-1}$. Indeed, since the unit cell of \mbox{\NCS{}} contains four ET molecules, every internal molecular vibration $\nu_i$ splits theoretically into four components. In addition, the mode degeneracy can be lifted by strong dimer interaction leading to additional modes.{\cite{Maksimuk2001, Swietlik1992}} As expected, modes at low temperature harden compared to room temperature. 


Let us discuss the origin of several modes.
The peaks at 2076 and 2110 cm$^{-1}$ in Fig.~\ref{Fig1} are related to distinct CN environments in the unit cell.\cite{kornelsenInfraredOpticalProperties1991} The strong intensity of the mode at 1476 cm$^{-1}$ suggests that this vibration is coupled to an electronic transition, leading to a resonant Raman scattering process. We have experimentally found that the intensity of the mode decreases at other wavelengths. Indeed, the laser wavelength (532 nm equivalent to 18797 cm$^{-1}$) is close to an electronic transition at 20000 cm$^{-1}$.\cite{zamboniRESONANTRAMANSCATTERING1989} This intense mode is assigned to the C=C stretching of the vibration of the central atoms of the ET molecule. 
The mode at 1510 cm$^{-1}$ is assigned to the a(a$_g$) vibration of the C=C ring. The peaks at 1421 and 1411 cm$^{-1}$ are assigned to CH$_2$ bending modes.\cite{zamboniRESONANTRAMANSCATTERING1989} It is interesting to note that the intensities of those peaks are strongly enhanced for the NPs. This could be a signature of a surface effect as ET molecules could be present at the surface of the NPs in a different arrangement than in the bulk and the ethylene groups less "bonded" to the crystal. The mode at 1041 cm$^{-1}$ is assigned to a C-C-H bending totally symmetric vibration and the mode at 778 cm$^{-1}$ is assigned to the C-S stretching of the ET molecule.\cite{kozlovAssignmentFundamentalVibrations1987}
To illustrate some vibrations, C=C stretching modes of the ET $\nu_2$, $\nu_3$ and $\nu_{27}$ are represented in Fig.~\ref{Fig2}.

\begin{figure}[h]
	\begin{center}
		\includegraphics[width=8.7cm]{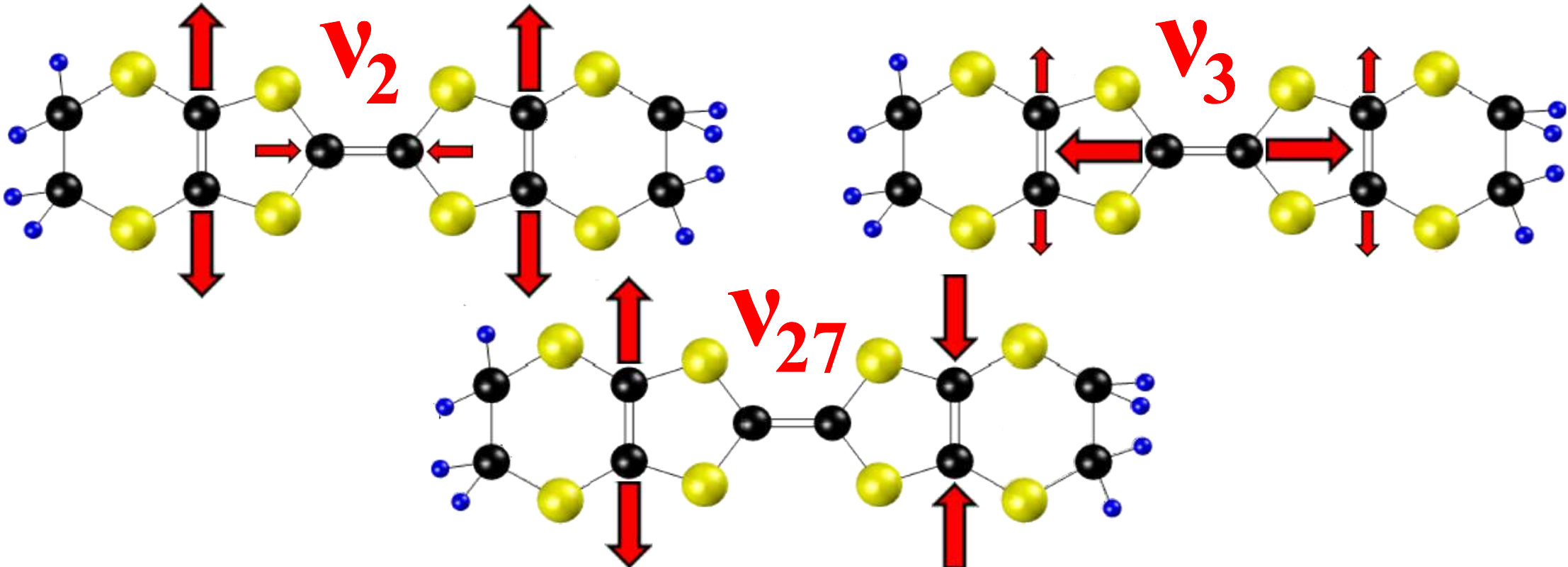}
		\caption{Atomic displacement vectors of the C=C stretching modes $\nu_2$, $\nu_3$ and $\nu_{27}$ (see Tab.~\ref{Table1}) in the ET molecule.\cite{Maksimuk2001}}
		\label{Fig2}
	\end{center}
\end{figure}


\subsection{Phonon modes in the low frequency region}

\NCS{} is monoclinic and crystallizes in the P2$_1$ space group with two formula units per cell (Z = 2). One unit cell is made of four ET molecules and two Cu(NCS)$_2$ zigzag polymeric chains. The ET molecules are paired and form two dimers in the cell. The unit cell contains 118 atoms, hence, the phonon spectrum consists of 354 phonon branches of which 288 pertain to internal modes of the ET molecules and 30 to internal modes of the polymeric anion layers. The remaining 36 branches consist of intermolecular motions of the ET molecules and the anion modes (including the three acoustic phonon branches). This calculation was derived according to Pedron {\it et al.}\cite{pedronPedronPhysicaCETBr1997Pdf1997}.

Due to the heavy mass of ET and [Cu(NCS)$_2$]$^-$, we expect intermolecular modes mostly in the low frequency spectral region. 

\begin{figure}[h]
	\begin{center}
		\includegraphics[width=8.7cm]{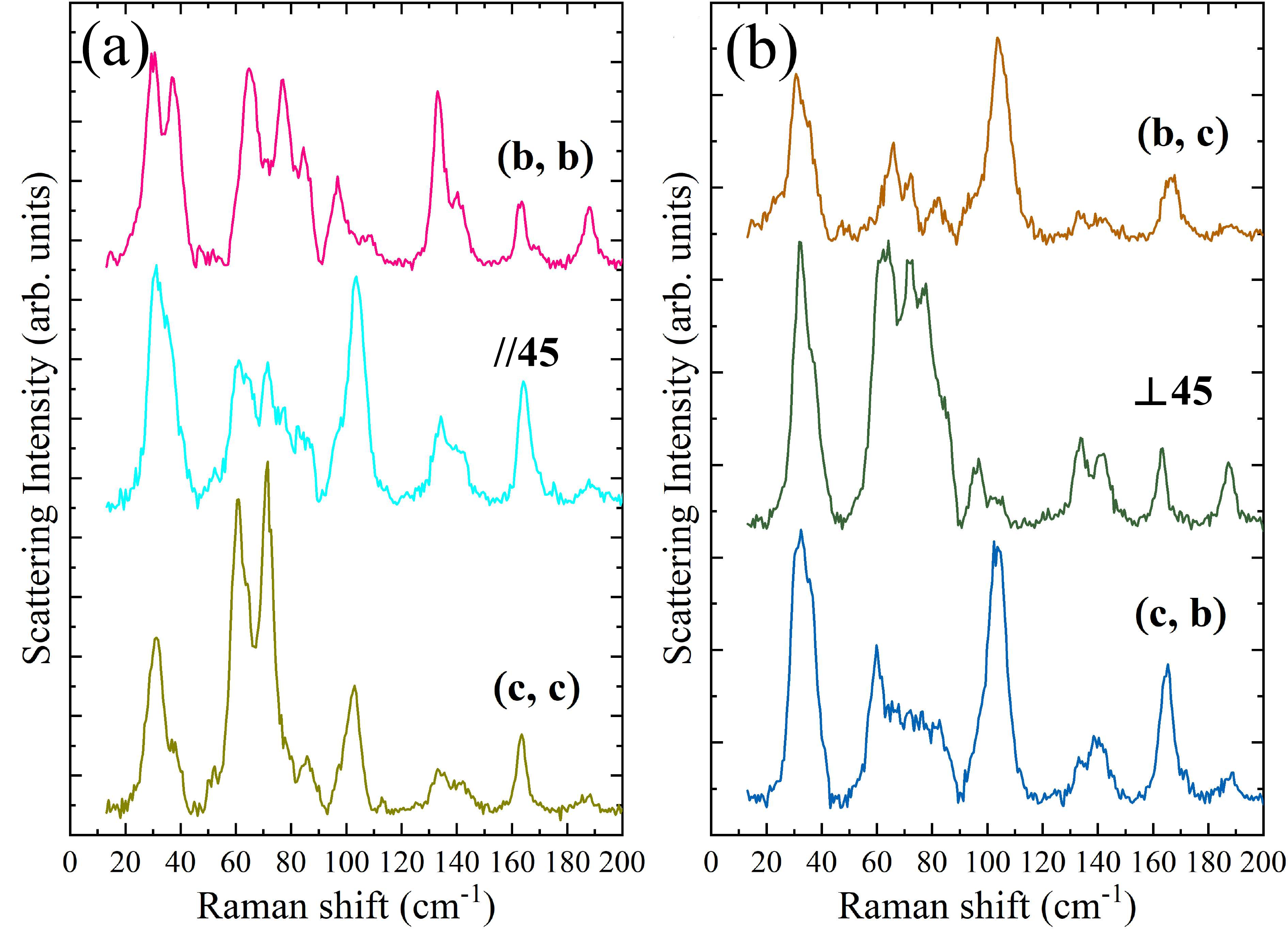}
		\caption{Polarized Raman spectra in the low frequency range measured at 15 K for the bulk using different polarization configurations as a function of the b and c axis of the crystal.}
		\label{Fig3}
	\end{center}
\end{figure}

\begin{table}
    \caption{Frequencies (in cm$^{-1}$) of Raman \NCS{} modes measured on the bulk (cf. Fig.~\ref{Fig3}) using several polarization configurations.}
    \begin{ruledtabular}
       \begin{tabular}{cccccc|cc}
            \multicolumn{6}{c|}{In our work} & \multicolumn{2}{c}{Pedron {\it et al.}\cite{pedronElectronphononCouplingBEDTTTF1999}}\\
            \cline{2-5} 
            (c,c) & (c,b) & //45 & $\perp$45 & (b,b) & (b,c) & (c,c) & (c,b) \Tstrut \\
            
            \hline
                  &       &       &       &       & 26.5  &       & 26.5 \Tstrut\\
            31.5  & 32.7  & 31.8  & 33.0  & 30.7  & 31.4  & 32.0  & 31.2        \\ 
            37.5  & 36.0  & 37.0  & 36.5  & 37.7  & 36.1  & 36.9  &             \\
            52.6  &       & 52.6  &       &       &       & 51.9  & 51.0        \\
            61.3  &       & 61.5  &       &       &       & 60.4  & 60.9        \\
                  & 63.3  & 64.7  & 63.4  &       &       & 64.3  & 64.8        \\
                  & 66.2  &       &       &       & 66.1  & 66.2  &             \\
            71.5  & 71.7  & 71.8  & 72.4  & 71.3  & 72.1  &       & 72.2        \\
                  &       & 77.1  & 77.2  & 77.4  &       & 74.6  & 75.0        \\
                  &       &       &       &       & 81.7  & 81.8  &             \\
                  & 83.2  &       &       & 84.9  &       &       & 83.1        \\
            90.0  &       &       &       &       &       & 90.5  & 90.5        \\
                  &       & 97.1  & 96.3  & 97.1  &       &       &             \\
            102.7 & 103.9 & 103.8 &       & 106.6 & 104.6 & 102.5 & 102.9       \\
            112.7 &       &       &       &       &       &       & 113.9       \\
            134.4 & 134.2 & 134.7 & 134.1 & 133.9 & 134.1 &       & 132.7       \\
            141.0 & 140.0 & 141.1 & 141.5 & 140.5 & 141.4 &       & 141.1       \\
            163.8 &       & 164.8 & 163.3 & 163.7 &       & 162.9 &             \\
                  & 165.3 &       &       & 168.3 & 166.8 &       & 164.7       \\
                  & 188.4 & 188.5 & 187.8 & 188.3 & 188.1 &       &             \\
        \end{tabular}
    \end{ruledtabular}
    \label{Table2}
\end{table}

Figure~\ref{Fig3} shows Raman spectra measured on the bulk for six different polarizations at 15 K in the low frequency region. The polarization configuration (i, j) indicates that the incident and scattered light are polarized along the i and j axis of the crystal, respectively. Special notation //45 ($\perp$45) indicates that the incident light is polarized at -45\degree with respect to the c axis and the scattered light at -45\degree(+45\degree) with respect to the c axis. In total, 20 phonon modes are observed in this range. The frequencies of the observed phonons are listed in Tab.~\ref{Table2} and compared to the measurements of Ref.~\onlinecite{pedronElectronphononCouplingBEDTTTF1999}. One notice that both measurements are in good agreement.

\begin{figure}[h]
    \begin{center}
    \includegraphics[width=8.7cm]{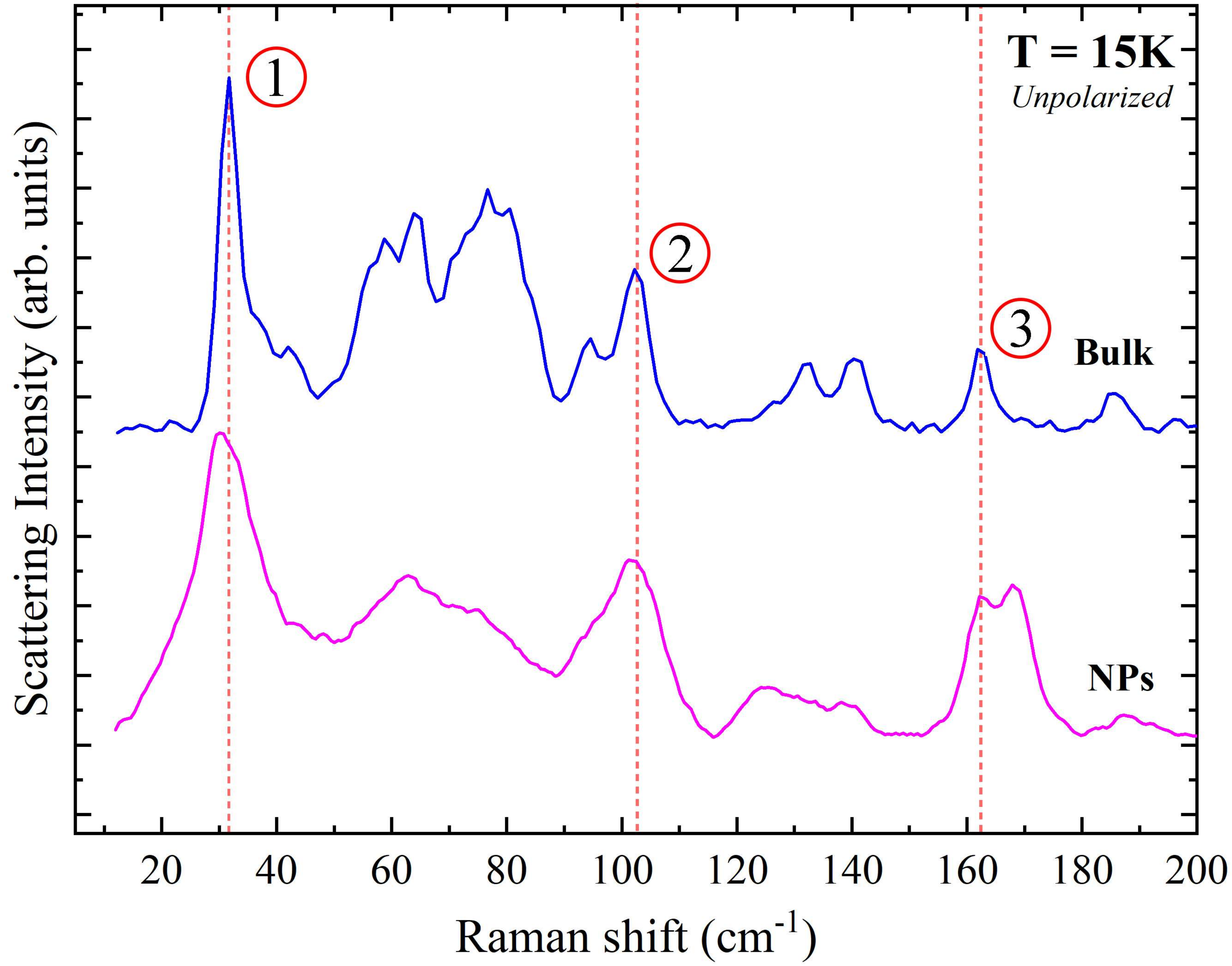}
    \caption{Unpolarized Raman spectra of bulk and NPs at 10 K in the low frequency range. Three phonons around 32 cm$^{-1}$, 103 cm$^{-1}$ and 163 cm$^{-1}$ are labelled 1, 2 and 3, respectively.}
    \label{Fig4}
    \end{center}
\end{figure}

Figure~\ref{Fig4} shows unpolarized Raman spectra measured on the bulk and NPs at 15 K in the range of 20 – 200 cm$^{-1}$. All the phonons measured in the bulk are found in the spectrum of NPs but with a larger width. Surface effect or grain boundaries in the NPs can explain a shorter phonon lifetime and therefore a larger width. 
In the low frequency regions, fewer Raman measurements have been made. J. E. Eldridge {\it et al.}  have detected phonons at 94, 118, 128 and 185 cm$^{-1}$ and they have associated them with anion lattice modes.\cite{Eldridge1997, Eldridge2002}  We have measured phonon modes at 97, 113, 134, and 188 cm$^{-1}$, and additional modes at 141 and 164 cm$^{-1}$. In the common range in energy, our measurements are in good agreement with this previous work.
In the next sections, we are following the temperature dependencies of three phonons around 32 cm$^{-1}$, 103 cm$^{-1}$ and 163 cm$^{-1}$ labeled 1, 2 and 3 in Fig.~\ref{Fig4}, respectively. 

\subsubsection{Bulk}

\begin{figure}[h]
	\begin{center}
		\includegraphics[width=8.7cm]{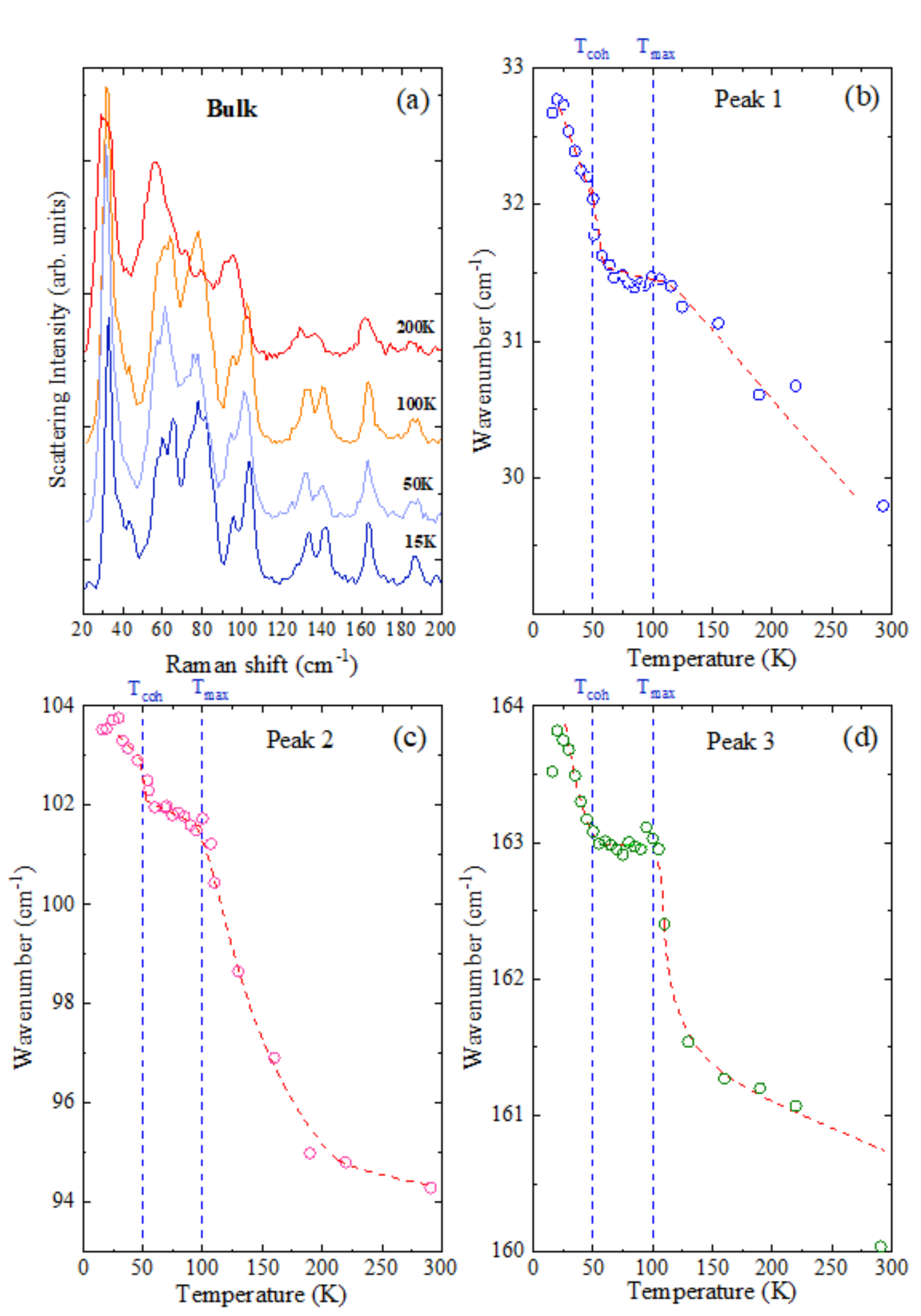}
		\caption{(a) Unpolarized Raman spectra measured on the bulk for several temperatures. (b-d) Frequencies of the 1, 2 and 3 phonon modes as a function of temperature, respectively. Experimental errors on phonons frequencies are around 0.2 cm$^{-1}$. The vertical lines corresponds to the transition temperatures $T_{coh}$ $\simeq$ 50 K and $T_{max}$ $\simeq$ 100 K.}
		\label{Fig5}
	\end{center}
\end{figure}

Figure~\ref{Fig5}(a) shows unpolarized Raman spectra measured on the bulk for several selected temperatures. Fig.~\ref{Fig5}(a), (b) and (c) presents the temperature evolution of labeled phonons 1, 2 and 3 (cf. Fig.~\ref{Fig4}). 
The frequencies of all three modes display essentially the same behavior, an increase down to $T_{max}$ $\simeq$ 100 K followed by a plateau (or a very small increase in the case of peak 2) between $T_{max}$ and $T_{coh}$ $\simeq$ 50 K before an increase at lower temperature. 
Note that the increase of phonon 2 and 3 is very abrupt and strongly non anharmonic close to 100 K.
Such a plateau is an unconventional behavior compared to the monotonous increase of the phonon frequency due to the thermal contraction of the lattice decreasing the temperature. 
X-ray measurements do not reveal any structural transition in this temperature range discouting the possibility that changes in the crystal structure could be responsible for the measured anomalies.\cite{Wolter2007}

The three phonon modes heavily involve the movement of the anions with respect to the ET molecules but also a complete bending of the molecules within the dimers.\cite{Dressel2016a, Dressel2016b} The terminal ethylene groups are probably taking part in these modes. The glassy transition at $T_g\simeq80$ K associated with the freezing of the terminal ethylene groups of the ET molecule,\cite{Muller2002} occurs in the middle of the temperature plateau. Therefore, the plateau observed in the phonon frequencies does not seem to be related to $T_g$. Moreover, if the glassy transition froze out the
relevant phonon modes, pronounced changes in intensity and shifts in the energy position of the related phonon
modes would be expected. 
This is supported by the work of Kuwata {\it et al.} suggesting that the decrease in the ethylene motion at $T_g$ does not contribute to the mechanism associated to $T_{max}$ even if the mechanism causing the $T_{max}$ anomalies is still under debate.\cite{Kuwata} The resistivity maximum was assigned to Wigner Mott scaling, for instance.\cite{Radonjic}

The behavior of the phonon modes between 50 K and 100 K is rather similar to that observed in resistivity measurements. The resistivity increases lowering the temperature from room temperature and reaches a broad maximum similar to a plateau around 100 K. The difference between our measurements and resistivity measurements appears at lower temperature. After the broad maximum, the resistivity falls rapidly decreasing the temperature.\cite{Analytis2006} This change in resistivity is understood in terms of the crossover from a bad metal at high temperatures to a Fermi liquid and the formation of quasiparticles.\cite{Milbrandt2013, Georges2013} Here the three phonon modes at low energy seem to be sensitive to the bad metal phase.

\subsubsection{Nanoparticles}

\begin{figure}[h]
	\begin{center}
		\includegraphics[width=8.7cm]{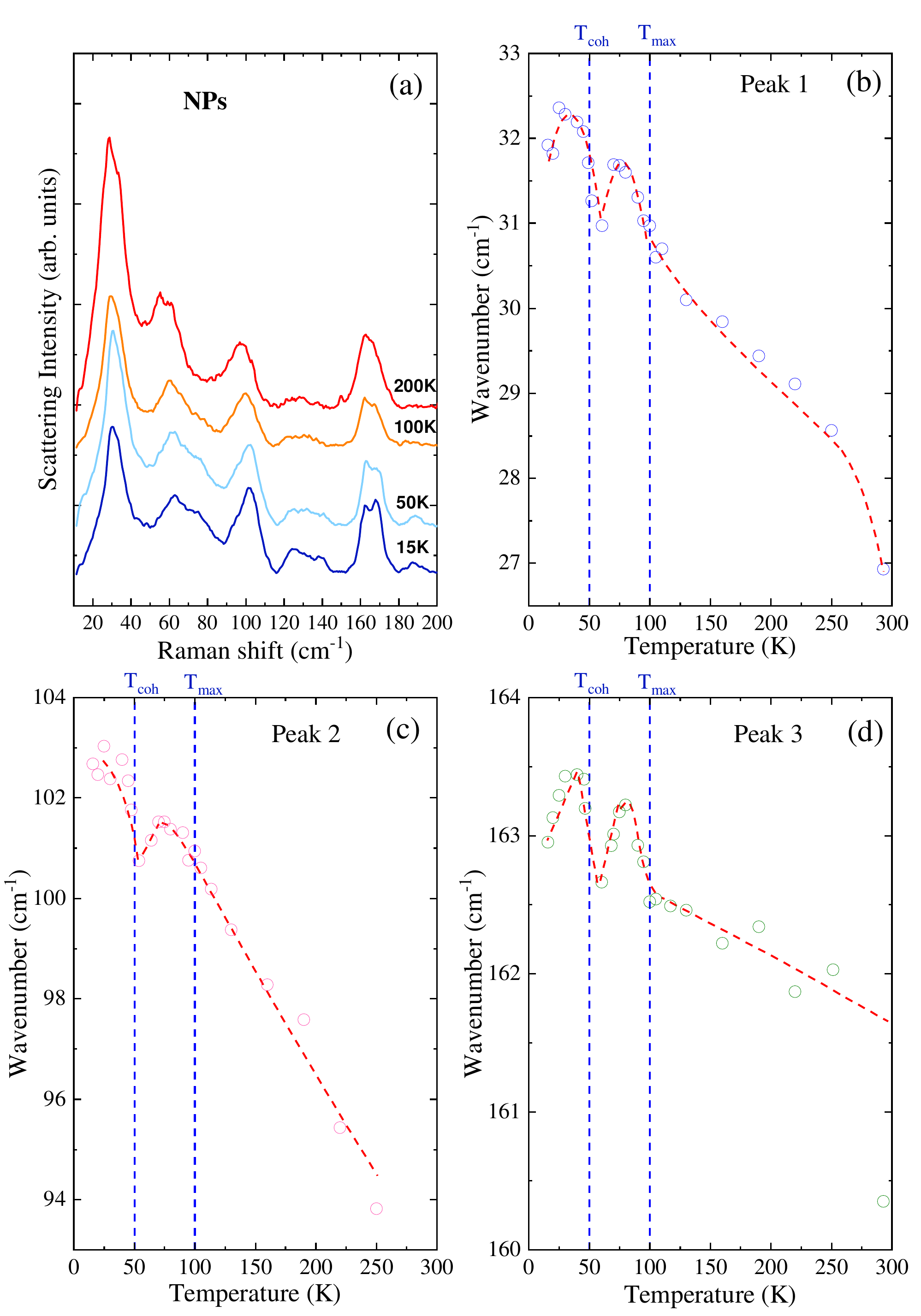}
		\caption{(a) Unpolarized Raman spectra of the NPs for several temperatures. (b-d) Frequencies of the 1, 2 and 3 phonon modes as a function of temperature, respectively. Experimental errors on phonons frequencies are around 0.2 cm$^{-1}$.}
		\label{Fig6}
	\end{center}
\end{figure}

Figure~\ref{Fig6} shows unpolarized Raman spectra of the NPs at selected temperatures between 10 and 250 K and the evolution of phonons 1, 2 and 3 (cf. Fig.~\ref{Fig4}). Note that no quadrupolar vibrational mode associated to the breathing mode of the NPs is observed.\cite{Duval1}
The frequencies of all three modes display essentially the same behavior 
although the peak 2 presents attenuated anomalies compared to the other two peaks.
In Fig.~\ref{Fig6}, we can see a deviation from the standard phonon behavior below 100 K and a double hump for the three phonons with a first minimum closed to $T_{coh}$ = 50 K and a small softening around $T_{FL}$ = 25K. Such behavior is different from the one observed in the bulk. Note that the amplitude of the minima are much higher than the experimental error on the frequency of the phonon modes (0.2 cm$^{-1}$). 

The softening of the phonon modes at 25 K might be related to the crossover to the pure Fermi liquid. This softening is also observed in the bulk (Fig.~\ref{Fig5}) but it is very small.

We can provide several approaches to explain the softening of the three phonon modes at $T_{coh}$ = 50 K, temperature associated to the crossover from a bad metal to a Fermi liquid.

From a theoretical point of view, the Hubbard-Holstein model using dynamical mean-field theory has been able to predict the crossover from a high temperature bad metal, characterized by the absence of quasiparticles, to a Fermi liquid, transition that occurs in \NCS  at $T_{coh}$.\cite{Merino2000HH}
Anomalies in the real part of the phonon self energy have been calculated from this model. Such anomalies induce the softening of the phonon frequencies at T$_\textrm{coh}$ = 50 K, the temperature associated to the quasiparticle formation, followed by the hardening of the frequencies at lower temperature.
The trend of the phonon frequencies in Fig.~\ref{Fig6} is similar to predicted modes with frequency around $U/2$ in Ref.~\onlinecite{Merino2000HH}. $U$ is the effective repulsion between two electrons on the same ET$_2$ dimer and $U/2$ is the location of the Hubbard bands in the spectral function.\cite{Merino2000HH} The value of $U/2$ has been calculated \cite{Scriven2009a,Scriven2009b} and reasonably estimated \cite{Dressel2007} around 1000 cm$^{-1}$. Our observations on the low frequency phonons located in frequency well below this value can not be explained by this model.
However, the Mott physic cannot be totally rule out. A strong enhancement of the low-energy spectral weight has been measured in the conductivity of related compounds as $\kappa$-(BEDT-TTF)$_2$Cu$_2$(CN)$_3$ close to the Mott transition.\cite{Pustogow} Note that, \NCS{} is located in the metallic side of the phase diagram of this family of compounds but it is not so far from the Mott transition. Similar to related organic compounds\cite{Dressel2007, Merino2008, Takenaka}, pronounced spectral weight shifts are expected at low frequencies upon the crossover between bad metal and Fermi liquid. 

From semiconducting to metallic phase, the electrodynamic response and the conductivity below 1000 cm$^{-1}$ are strongly modified.  As a consequence, the electronic background which is a function of the polarizability and the coupling to the phonons should change.

The opening of a pseudogap below $T_\textrm{coh} \simeq$ 50 K or a rapid decay in the spin fluctuations associated with the formation of quasiparticles has been proposed to interpret nuclear magnetic resonance experiments where a sharp maximum in the spin fluctuations is observed at $T_\textrm{coh}$ $\simeq$ 50 K.\cite{Kawamoto95,Kawamoto95B, Powell2011} Both possibilities could be at the origin of the softening of the three phonon modes at $T_{coh}$ = 50 K. 

To distinguish them, a first argument is given by the expected signature of a pseudogap in Raman spectra. A pseudogap opening is characterized by a modification of spectral weight in the electronic background from above to below the pseudogap temperature. We did not observe such a behaviour in the electronic background of our spectra. 
So far, no microscopic model has been developed to describe the phonon behavior in the presence of a pseudogap, however one knows that the opening of a gap can be associated to phonon softening. In high temperature superconductors, phonons with an energy below the energy of the pair-breaking peak soften, whereas they harden if their energies are above the gap.\cite{Zeyher1990,Bakr2009}
In our experiment, the phonon 3 in Fig.~\ref{Fig6} softens significantly below $T_\textrm{coh}$. 
To be consistent with a possible gap opening, the energy of this gap must be equal to or higher than the energy of the highest phonon that is softening. This situation corresponds to an opening of a pseudogap above 163 cm$^{-1}$ = 237 K.
However this energy is higher than $T_\textrm{coh}$. This means that the softening of the three phonon modes at $T_\textrm{coh}$ has little chance from being related to the opening of a pseudogap.
Remain the rapid decay in the spin fluctuations associated with the formation of quasiparticles as a possibility to explain our measurements below 50 K in NPs. Under such a hypothesis, the spin-phonon coupling might be the link between the observed phonon anomalies and the spin fluctuations. 
Although we cannot accurately determine the origin of these phonon anomalies, the measured phonon softenings at $T_\textrm{coh}$ show that these phonon modes are connected to the transition from the bad metal to the Fermi liquid phase. 

One question remains. Why are theses anomalies observed in the NPs and not in the bulk ? 
NPs differ qualitatively from the bulk due to their larger surface-to-volume ratios. Knowing that our NPs are 28 nm diameter-large, the surfaces effects on the bulk properties can no longer be overlooked. Reduced symmetry at the surface increases surface anisotropy due to factors including broken exchange bonds and interatomic distance variation. This could be at the origin of the greater sensitivity of the phonons of NPs to spin fluctuations.\\

\section{Conclusion}

In summary, we have investigated the organic compound \NCS{} in the form of bulk and NPs by Raman light scattering and observed anomalies in the low frequency phonon energies as a function of temperature. In the bulk, the changes in peak energy between 100 and 50 K are attributed to the signature of the bad metal. For NPs, the anomalies around and below $T_\textrm{coh}$ $\simeq$ 50 K are related to the transition from the bad metal to the Fermi liquid and might be associated to spin fluctuations and their rapid decay.
  

\end{document}